\documentclass[twocolumn,pra,aps,amssymb,amsfonts]{revtex4}
\usepackage{times}

\newcommand{\trace}{\mathop{\rm Tr}\nolimits}

\newcommand{\qed}{\hfill$\square$\par\vskip24pt}

\newcommand{\cH}{{\cal H}}

\newcommand{\id}{\openone}

\newcommand{\be}{\begin{equation}}
\newcommand{\ee}{\end{equation}}
\newcommand{\bea}{\begin{eqnarray}}
\newcommand{\eea}{\end{eqnarray}}
\newcommand{\beas}{\begin{eqnarray*}}
\newcommand{\eeas}{\end{eqnarray*}}

\newtheorem{theorem}{Theorem}
\newtheorem{lemma}{Lemma}
\newtheorem{corollary}{Corollary}

\begin{document}
\title{Subadditivity of $q$-entropies for $q>1$}
\author{Koenraad M.R. Audenaert}
\email{k.audenaert@imperial.ac.uk}
\affiliation{Institute for Mathematical Sciences, Imperial College London,
53 Princes Gate, London SW7 2PG, UK}

\date{\today}

\begin{abstract}
I prove a basic inequality for Schatten $q$-norms of quantum states
on a finite-dimensional bipartite Hilbert space $\cH_1\otimes \cH_2$:
$1+||\rho||_{q} \ge ||\trace_1\rho||_{q} + ||\trace_2\rho||_{q}$.
This leads to a proof---in the finite dimensional case---of
Raggio's conjecture (G.A.~Raggio, J.\ Math.\ Phys.\ \textbf{36}, 4785--4791 (1995))
that the $q$-entropies $S_q(\rho)=(1-\trace[\rho^q])/(q-1)$
are subadditive for $q > 1$; that is, for any state $\rho$ ,
$S_q(\rho)$ is not greater than the sum of the $S_q$ of its reductions,
$S_q(\rho) \le S_q(\trace_1\rho)+S_q(\trace_2\rho)$.
\end{abstract}
\maketitle

In this Note I obtain an inequality relating the Schatten $q$-norm \cite{bhatia} of a quantum state
on a finite-dimensional bipartite Hilbert space $\cH_1\otimes \cH_2$,
to the $q$-norms of its reductions to $\cH_1$ and $\cH_2$.
These reductions are given by the partial traces
$\rho_1=\trace_2\rho$ and $\rho_2=\trace_1\rho$.
Partial traces are linear operations defined by $\trace_1: X\otimes Y\mapsto \trace[X]Y$ and
$\trace_2: X\otimes Y\mapsto \trace[Y]X$, for general square matrices $X$ and $Y$.
The Schatten $q$-norms are non-commutative generalisations of the familiar $\ell_q$-norms. For the
special case of positive semi-definite matrices (including states), they are defined as \cite{bhatia}
$$
||A||_q := (\trace[A^q])^{1/q}.
$$
I obtain the following Theorem:
\begin{theorem}
For any bipartite state $\rho$ on a Hilbert space $\cH_1\otimes \cH_2$, the inequality
\be\label{eq:main}
1+||\rho||_{q} \ge ||\trace_1\rho||_{q} + ||\trace_2\rho||_{q}
\ee
holds for $q>1$.
\end{theorem}
Equality holds, for example, for product states $\rho=\rho_1\otimes\rho_2$
where at least one of the factors $\rho_i$ is pure (i.e.\ has rank 1).

A straightforward argument exploiting this Theorem then leads to a proof of subadditivity of the
so-called $q$-entropies, for $q>1$.
These $q$-entropies are defined as \cite{tsallis}
\be
S_q(\rho)=(1-\trace[\rho^q])/(q-1).
\ee
In the limit $q\to1$, $S_q$ reduces to the von Neumann entropy $S(\rho):=\trace[\rho\log\rho]$,
which already was known to be subadditive \cite{wehrl}.

\begin{theorem}
For any bipartite state $\rho$ on a Hilbert space $\cH_1\otimes \cH_2$, the inequality
\be\label{eq:sub}
S_q(\rho) \le S_q(\trace_1\rho)+S_q(\trace_2\rho)
\ee
holds for $q>1$.
\end{theorem}
This second Theorem proves a conjecture by GA Raggio from \cite{raggio} for
finite-dimensional bipartite quantum states.
For classical finite-dimensional states (i.e.\ 2-variate finite-dimensional probability distributions)
this was proven by Raggio in \cite{raggio}. Still in the classical case, but for continuous distributions,
the conjecture was proven for $1\le q\le 2$ and refuted for $q>2$ by Bercovici and
Van Gucht \cite{bercovici}.
For $q<1$, the $q$-entropies are superadditive on product states.
For general states they are neither subadditive nor superadditive \cite{raggio}.

\bigskip

I now present the proofs of the above Theorems.
First, let $(x)_+$ denote the function $x\mapsto\max(0,x)$.
Similarly, for a Hermitian matrix $X$, let $X_+$ denote the positive part of $X$,
which is obtained by replacing each one of the eigenvalues $\lambda_i$ of $X$ by $(\lambda_i)_+$.
Then we have the following Lemma for finite-dimensional non-negative vectors and a subsequent Corollary
generalising it to positive semi-definite matrices.
\begin{lemma}
Let $q>1$.
Let $x=(x_1,x_2,\ldots)$ and $y=(y_1,y_2,\ldots)$ be two non-negative real vectors, normalised according to
the $\ell_q$ norm, i.e.\ $||x||_q=||y||_q=1$.
Then the inequality
\be
\sum_{i,j} ((x_i+y_j-1)_+)^q \le 1
\ee
holds.
\end{lemma}
\textit{Proof.}
Let $x$ and $y$ be the vectors of the Lemma.
Consider the function
$$
f(a):=||(y+a-1)_+ ||_q = \Big(\sum_j ((y_j+a-1)_+)^q\Big)^{1/q}.
$$
This function is convex in $a$ because the $\ell_q$ norm for non-negative real vectors is
convex and monotonously increasing in each of the vector's entries, and because the function $a\mapsto (b+a-1)_+$
is convex for any real value of $b$.
Furthermore, its values in $a=0$ and $a=1$ are 0 and 1 respectively, since $0\le y_j\le 1$ and $||y||_q=1$.
Therefore, the inequality $f(a)\le a$ holds for $0\le a\le 1$.
As each of the $x_i$ obeys $0\le x_i\le 1$, we have
\beas
\sum_{i,j} ((x_i+y_j-1)_+)^q &=& \sum_i f(x_i)^q \\
&\le& \sum_i x_i^q =1,
\eeas
which is what we needed to prove.
\qed

Note that the above proof is essentially discrete (finite or countably infinite) and does not work
in the continuous case.
Furthermore, the Lemma itself cannot even be true in the continuous case because the proof
presented below would then also go trough for continuous distributions, which cannot be
since subadditivity for continuous distributions does not hold for $q>2$ \cite{bercovici}.
The essential point where the proof of the Lemma fails in the continuous case
is that for non-negative functions $f$
on a probability space $\Omega$ with probability measure $\mu$,
individual values of $f$ can be larger than its $\ell_q$ norm $||f||_q:=(\int_\Omega d\mu f^q)^{1/q}$
(so that $||f||_q=1$ does not imply $f\le 1$), unless $\mu$ is a counting measure,
such as in the finite-dimensional case considered by Raggio (and here).

\bigskip

The Lemma can be reformulated in terms of positive semi-definite matrices.
\begin{corollary}
Let $X$ and $Y$ be positive semi-definite matrices with $||X||_q=||Y||_q=1$.
Then the inequality
\be
||(X\otimes\id + \id\otimes Y-\id)_+||_q \le 1
\ee
holds for $q>1$.
\end{corollary}
\textit{Proof.}
Since $X$ and $Y$ are positive semi-definite, they can be unitarily diagonalised. Let the
obtained diagonal matrices have diagonal entries $x_i$ and $y_j$ respectively. These are
non-negative real numbers satisfying the conditions of the Lemma.
Since the Schatten $q$-norm of a positive matrix is equal to the $\ell_q$-norm of its eigenvalues, we get
$$
||(X\otimes\id + \id\otimes Y-\id)_+||_q = || (x\otimes e + e\otimes y-e)_+ ||_q,
$$
where $e$ is shorthand for an all-ones vector $(1,\ldots,1)$ of appropriate dimension.
The entries of the vector appearing in the right-hand side are exactly $(x_i + y_j-1)$, so that
by the Lemma the right-hand side is upper bounded by 1.
\qed

\textit{Proof of Theorem 1.}
A simple consequence of the Corollary is that for all $X$ and $Y$ of Schatten $q$-norm equal to 1,
a positive semi-definite matrix $Z$ of Schatten $q$-norm 1 exists that obeys the matrix inequality
$$
Z\ge X\otimes\id + \id\otimes Y-\id.
$$
Indeed, the positive part $H_+$ of any Hermitian matrix $H$ obeys $H_+\ge H$. In particular, thus,
$$
(X\otimes\id + \id\otimes Y-\id)_+ \ge X\otimes\id + \id\otimes Y-\id.
$$
By the Corollary, the $q$-norm of this positive part is upper bounded by 1. It is therefore
possible to add a positive matrix to $(X\otimes\id + \id\otimes Y-\id)_+$ and obtain a positive matrix $Z$
with $q$-norm exactly 1. This follows immediately from Weyl's monotonicity principle \cite{bhatia}:
for $A,B\ge0$, $||A+B||_q\ge||A||_q$.

For these $X$, $Y$ and $Z$ we then have, for any normalised state $\rho$,
\beas
\trace[Z\rho]+1 &=& \trace[(Z+\id)\rho] \\
&\ge& \trace[(X\otimes\id + \id\otimes Y)\rho] \\
&=& \trace[X\rho_2 + Y\rho_1].
\eeas
I will now exploit the fact that the Schatten $q$-norms have a dual representation \cite{bhatia}.
Let $q'$ be such that $1/q+1/q'=1$.
Then for positive semi-definite $A$, one has
$$
||A||_{q'} := \max_{B\ge 0} \{\trace[AB]: ||B||_{q}\le1\}.
$$
Let us now choose $X$ and $Y$ in such a way that $\trace[X\rho_2] = ||\rho_2||_{q'}$ and $\trace[Y\rho_1]=||\rho_1||_{q'}$.
In words, we choose $X$ and $Y$ to be the optimal variational arguments ($B$) in the dual representation
of $||\rho_2||_{q'}$ and $||\rho_1||_{q'}$, respectively.
The matrix $Z$ corresponding to these $X$ and $Y$ will in general be suboptimal in the dual representation of $||\rho||_{q'}$,
so that $\trace[Z\rho]\le ||\rho||_{q'}$ holds.
After dropping primes we obtain the inequality (\ref{eq:main}).
\qed

\bigskip

\textit{Proof of Theorem 2.}
To obtain inequality (\ref{eq:sub}), note that ineq.\ (\ref{eq:main}) can be written as a 1-norm inequality
for the positive 2-vectors $u:=(1,||\rho||_{q})$ and $v:=(||\rho_1||_{q},||\rho_2||_{q})$, namely as
$||u||_1 \ge ||v||_1$.
Since $\rho$ and its reductions are normalised states, their $q$-norms are upper bounded by 1, so that
the inequality $||u||_\infty \ge ||v||_\infty$ follows trivially.
As a consequence, the vector $u$ weakly majorises $v$. By Ky Fan's dominance theorem \cite{bhatia},
we get $|||u||| \ge |||v|||$ for any symmetric norm, and for the $\ell_q$ norm in particular.
Thus we obtain
\be
1+||\rho||_{q}^q \ge ||\rho_1||_{q}^q+||\rho_2||_{q}^q,
\ee
which is equivalent to inequality (\ref{eq:sub}).
\qed

\begin{acknowledgments}
This work is part of the QIP-IRC (\texttt{www.qipirc.org}), supported by
EPSRC (GR/S82176/0). The author is supported by the Institute of Mathematical Sciences,
Imperial College London.
\end{acknowledgments}


\begin{thebibliography}{9}
\bibitem{bercovici} H.~Bercovici and D.~Van~Gucht, Math.\ Ineq.\ and Appl.\ \textbf{8}, 743--748 (2005).
\bibitem{bhatia} R.\ Bhatia, \textit{Matrix Analysis}, Springer, Heidelberg (1997).
\bibitem{raggio} G.A.~Raggio, J.\ Math.\ Phys.\ \textbf{36}, 4785--4791 (1995).
\bibitem{tsallis} C.~Tsallis, J.\ Stat.\ Phys.\ \textbf{52}, 479--487 (1988).
\bibitem{wehrl} A.~Wehrl, Rev.\ Mod.\ Phys.\ \textbf{50}, 221--260 (1978).
\end{thebibliography}
\end{document}